\documentclass[preprint,12pt, a4paper]{elsarticle}
\usepackage[utf8]{inputenc}
\usepackage[margin=1.5cm]{geometry}
\usepackage{titlesec}
\usepackage{tabu}
\usepackage{enumitem}
\usepackage{amssymb}
\usepackage{xcolor}
\newlist{selectlist}{itemize}{2}
\setlist[selectlist]{label=$\square$,leftmargin=*,noitemsep,topsep=0pt}

\usepackage{lmodern}
\usepackage{graphicx}
\usepackage{hyperref}
\usepackage{float}
\hypersetup{
    colorlinks=true,
    linkcolor=blue,
    filecolor=magenta,      
    urlcolor=blue,
}
 
\titleformat{\section}[block]{\hspace{1em}\bfseries}{\thesection.}{0.5em}{} 
\titleformat{\subsection}[block]{\hspace{1em}}{\thesubsection}{0.5em}{}

\usepackage{lineno}

\begin{document}

\noindent
\textbf{\textit{Grafana plugin for visualising vote based consensus mechanisms, and network P2P overlay networks}}
\vskip0.5cm
\noindent
\textbf{\textit{Daniil Baldouski, University of Primorska, d.baldovskiy@mail.ru}}\\
\textbf{\textit{Aleksandar Tošić, University of Primorska, aleksandar.tosic@upr.si}}\\

\noindent
\textbf{Abstract}\\
In this paper, we present a plugin for visualising vote based consensus mechanisms primarily aimed to help engineers understand and debug blockchain and distributed ledger protocols. Both tools are built as Grafana plugins and make no assumptions on the data storage implementation. The plugins can be configured via Grafana plugin configuration interface to fit the specifics of the protocol implementation.
\vskip0.5cm

\noindent
\textbf{Keywords}\\
\textit{Grafana, Visualisation, Consensus mechanism, Blockchain protocols, P2P, Overlay network}
\vskip0.5cm
\newpage
\noindent
\textbf{Code metadata}\\

\noindent
\begin{tabular}{|l|p{6.5cm}|p{9.5cm}|}
\hline
\textbf{Nr.} & \textbf{Code metadata description} & \textbf{Please fill in this column} \\
\hline
C1 & Current code version & \textit{
v 1.0.2, v 1.0.0} \\
\hline
C2 & Permanent link to code/repository used for this code version &
\textit{
\underline{\url{https://gitlab.com/rentalker/consensus-visualization-plugin}},
\underline{\url{https://gitlab.com/rentalker/topology-visualization-plugin}}
} \\
\hline
C3  & Permanent link to Reproducible Capsule & \textit{
\underline{\url{https://gitlab.com/rentalker/grafana-docker-image}}
} \\
\hline
C4 & Legal Code License   & \textit{MIT} \\ 
\hline
C5 & Code versioning system used & \textit{git} \\
\hline
C6 & Software code languages, tools, and services used & \textit{TypeScript}\\
\hline
C7 & Compilation requirements, operating environments \& dependencies &  \textit{Node.js v14.0.0 or later, Yarn for compilation}\\
\hline
C8 & If available Link to developer documentation/manual & \textit{
\underline{\url{https://gitlab.com/rentalker/consensus-visualization-plugin/-/blob/master/README.md}},
\underline{\url{https://gitlab.com/rentalker/topology-visualization-plugin/-/blob/master/README.md}}
} \\
\hline
C9 & Support email for questions & \textit{d.baldovskiy@mail.ru} \\
\hline
\end{tabular}\\
\vskip0.5cm
\noindent
\section{Visualising vote based consensus in distributed systems}
In the past decade, decentralized systems have been increasingly gaining more attention. Much of the attention arguably comes from both financial, and sociological acceptance, and adoption of blockchain technology. One of the frontiers has been the design of new consensus protocols, topology optimisation in these peer to peer(P2P) networks, and gossip protocol design. Analogue to agent based systems, transitioning from the design to implementation is a difficult task. This is due to the inherent nature of such systems, where nodes or actors within the system only have a local view of the system with very little guarantees on availability of data. Additionally, such systems often offer no guarantees of a system wide time synchronisation.

Debugging distributed systems is a notoriously difficult task. Most standard debugging techniques such as logging, and profiling fall short as they cannot capture the global view of the system. There have been many attempts at building better tools and frameworks that tackled debugging distributed systems. One of the studied approaches is to give engineers tools to visualise the performance of a distributed system, and observe the discrepancies between the expected behaviour, and actual. The early attempts at visualising distributed was aimed at providing students insight into the behaviour of algorithm through animation (eg. \cite{socha1988voyeur},\cite{roman1992pavane}). \\
\noindent
\subsection{Common debugging methods}
Distributed and decentralized systems are difficult to debug as developers are working on the third layer (L3). Which includes L1(code level bugs), issues with concurrency on L2 (individual run-time), and finally the third dimension for potential bugs arising from the message exchange between nodes. In general, it is often hard to capture the state in a distributed system as debuggers can not be attached to all node's run-times. Additionally, it is often difficult to reproduce errors when they are inherently stochastic.

Common debugging techniques:
\begin{itemize}
    \item \underline{Logging} is the most common debugging method for all three layers. However, in distributed systems it is important to aggregate logs, and analyze them in as a time series. Additionally, aggregating distributed logs assumes the system has some method of clock synchronization protocol. The aggregation can be done with specific tools for log collection such as Prometheus, Logstash, etc. 
    \item \underline{Remote debugging} is a technique where a locally running debugger is connected to a remote node in the distributed system. This allows developers to use the same features as if they were debugging locally. However, it is difficult to determine which remote node should be debugged. Additionally, in case of Byzantine behaviour due to network faults connecting the debugger could fail.
    \item \underline{Distributed deterministic simulation and replay} is a technique that attempts to address the issues of reproducibility in distributed systems. The technique suggests implementing an additional layer that abstracts the underlying hardware and the network interfaces to allow for an exact replay of all the state changes and messages exchanged between nodes. Tools such as FoundationDB (\url{https://apple.github.io/foundationdb/testing.html}) or even custom systems built on containerisation software
    \item \underline{Visualisation and time series analysis} attempts at capturing the state of the system, and all the nodes by visualising the collected logs. Tools like Prometheus \cite{turnbull2018monitoring} and Grafana \cite{chakraborty2021grafana} are used extensively.
    
\end{itemize}
\noindent
\subsection{Consensus protocol, and consensus roles}
The aim of consensus protocols is to prevent a single entity in the system to change the global state. In blockchain protocol, state is snapshot in blocks, every block added to the chain the system transitions to a new state. From a top level view, each node within the has perceived relative truth, through execution of the consensus protocol, nodes need to reach an absolute truth from relative truths. In vote based consensus protocols, nodes vote for proposed blocks, a block is accepted by the protocol if it gathers sufficient amount of votes. However in public permission-less networks, it is difficult to have all nodes participate in the protocol for each block. The issues are scalability related, where congestion in P2P networks is hardly avoidable since votes need to be broadcast through the network. To circumvent this, state of the art blockchain protocols \cite{buterin2017casper} regularly shuffle nodes, and select a subset of them to vote. As long as the selected subset of nodes is large enough (representative sample), and nodes were chosen at random, the rest of the network can accept the block provided it passed the vote.

This research was inspired from implementation issues of a vote based consensus mechanism similar to aforementioned protocols \cite{kiayias2017ouroboros}. The main difference is that randomness is provided with verifiable delay functions(VDF) \cite{boneh2018verifiable}. The proof from a VDF is an entropy pool for a random number generator(RNG). Nodes are then able to shuffle the set of nodes participating in the consensus refereed to as the validator set.
Each block, nodes are assigned random roles and execute sub-protocols depending on their role.
\begin{itemize}
    \item \underline{Validators} are generic roles for all nodes participating in the consensus protocol. Their role is to validate the block by verifying the signatures, and transactions within the block.
    \item \underline{Committee} members are a subset of nodes that is entrusted to vote for the proposed block. Their votes are BLS12-381 \cite{wahby2019fast} signatures that get aggregated into a single signature while maintaining variability.
    \item The role of the \underline{block producer} is to prepare a candidate block, broadcast it to the committee members and prepare to receive aggregated votes, verify them and broadcast the final block with the aggregated votes to validators.
\end{itemize}
For each round, these roles get shuffled in a random fashion. However, the protocol also has fail saves to achieve liveliness through skip blocks in case nodes fail to perform their roles.

\section{Visualisation Tools}
\begin{figure}[!ht]
\centering
\includegraphics[scale=0.7]{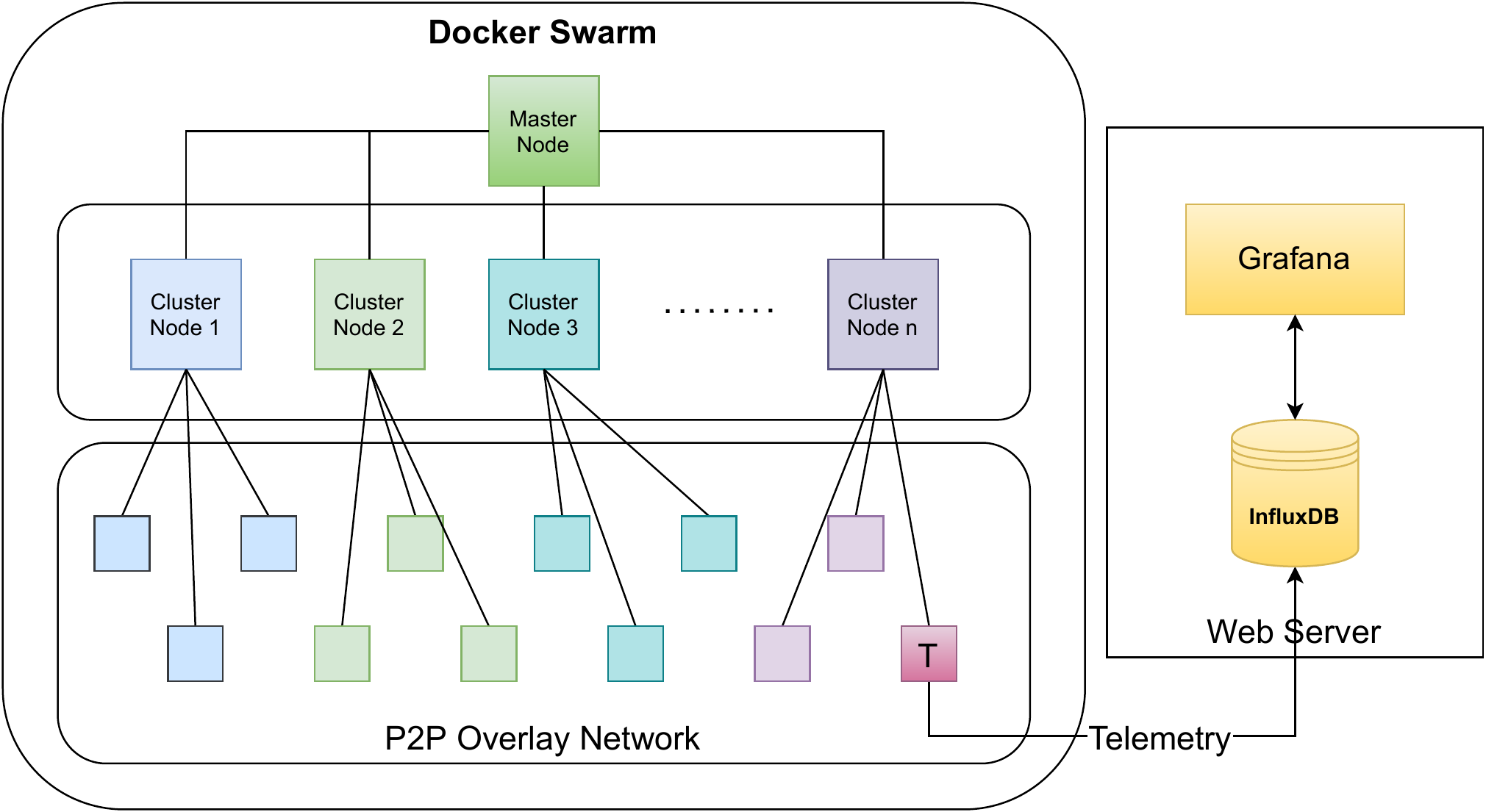}
 \caption{System architecture}
 \label{fig:architecture}
\end{figure}

We have developed two plugins that extend the functionality of Grafana. Both plugins were developed as React components and their life-cycle is managed by Grafana.
Figure \ref{fig:architecture} outlines the architecture used in production. A server running a database instance (preferably time series i.e. InfluxDb), and the Grafana platform. Depending on the underlying blockchain implementation, nodes can insert their telemetry directly to the database, or if possible have an archive node gather telemetry from nodes, and report them. In this example, a cluster was used to run multiple nodes. A coordinating node is responsible for maintaining an overlay network and serving the nodes within the overlay with a DHCP, DNS, and routing. Nodes are packed within docker containers and submitted to the coordinator, which uses built in load balancing and distributes them to other cluster nodes. 

\begin{figure*}[!ht]
\centering
\includegraphics[scale=0.35]{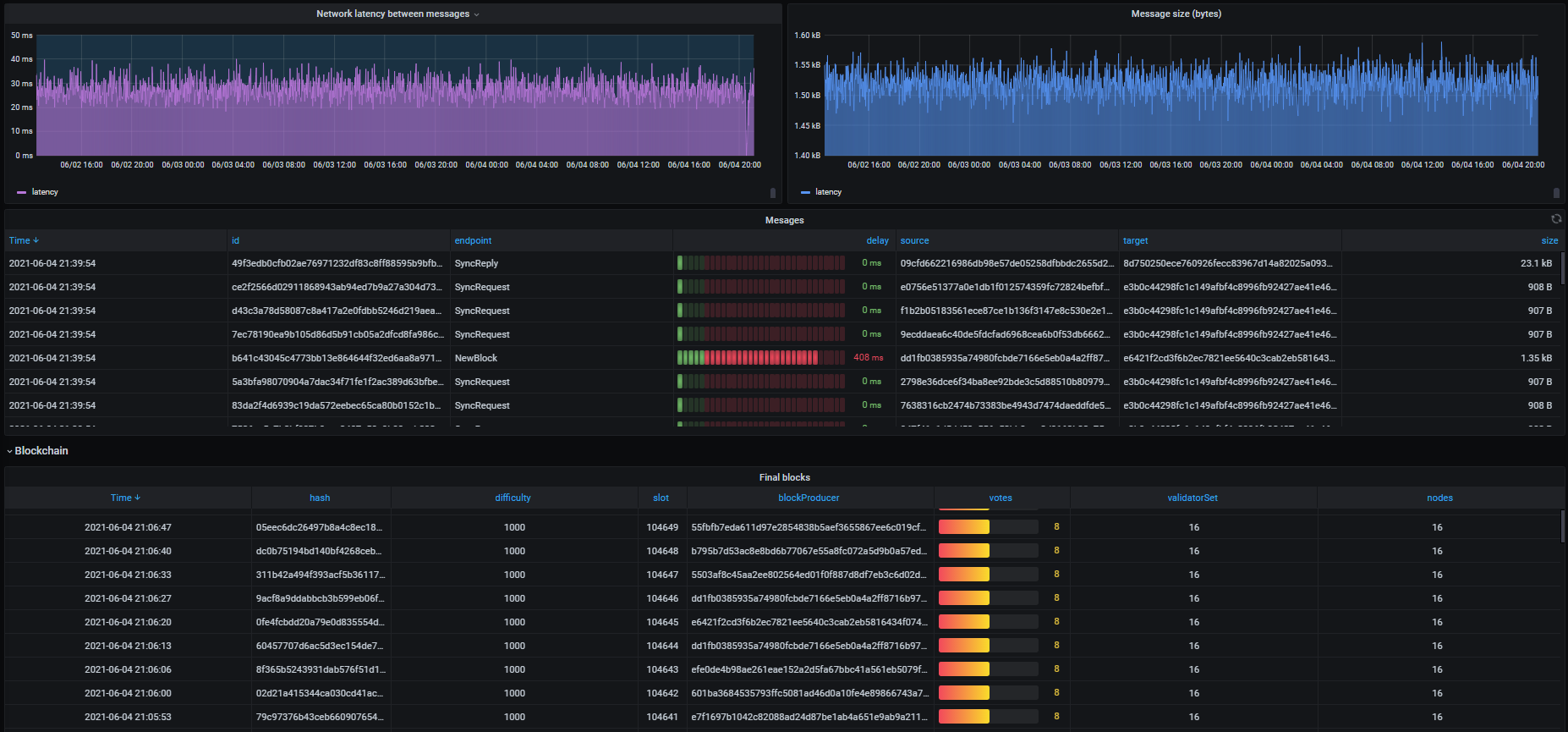}
 \caption{Part of the Grafana dashboard used by developers to gain insight into a running PoS based blockchain network}
 \label{fig:screenshot}
\end{figure*}

The telemetry inserted is timestamped to create a time series stream of data that is consumed by Grafana. Figure \ref{fig:screenshot} shows a small part of the dashboard created within Grafana using the built in plugins for visualisations. These visualisations are time series data of a running blockchain showing telemetry reported by the nodes.

\subsection{Network Plugin}
P2P networks propagate information using gossip protocols. There are many variations of the general and implementation specifics but in general the family of protocols aims at gossiping the fact that new information is available in the network. Should a node hear about the gossip, and require the information it will contact neighbouring nodes asking for the data. In general, gossip protocols make no assumptions about the topology of the overlay network. However, with structured networks, the information exchange can be made much more efficient. The observed blockchain implementation utilized a semi structured network topology for propagating consensus based information. This is made possible by using a seed string shared between nodes that is used for pseudo-random role election every block. Using the seed, nodes self-elect into roles without the need to communicate. However, when performing roles, committee members must attest to the candidate block produced by the block producer. The seeded random is therefore also used to cluster the network using a k-means algorithm. The clustering is again performed by each node locally. The shared seed guarantees that nodes will produce the same topology, which is then used to efficiently propagate attestations to the block producer.

\begin{figure*}[!ht]
\centering
\includegraphics[scale=0.5]{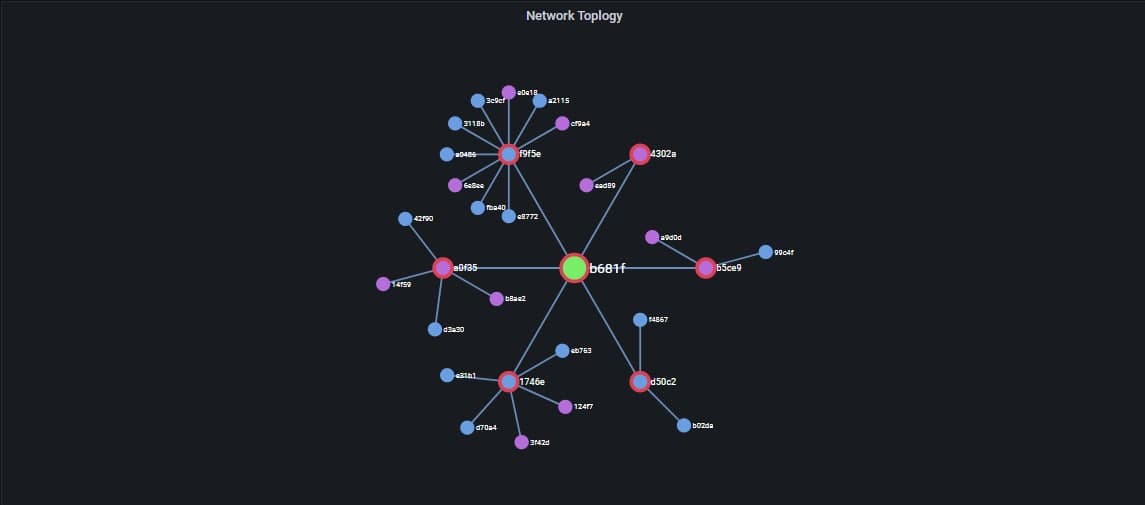}

 \caption{Network topology plugin visualising a test network of 30 nodes in real time.}
 \label{fig:topology}
\end{figure*}

The network topology hence changes every slot. The plugin aims to visualize the changes in the network topology by drawing nodes, and their cluster representatives. Additionally, the consensus roles for each node are indicated with the vertex color. Figure \ref{fig:topology} shows the network plugin rendering a test network of 30 nodes real-time. The node in the center coloured green is the elected block producer for the current slot, nodes surrounded by the red stroke are cluster representatives, the rest of the nodes are coloured based on their role in the current slot, and connected to their cluster representative respectively.

\subsection{Consensus Plugin}
\begin{figure}[H]
\centering
\includegraphics[scale=0.5]{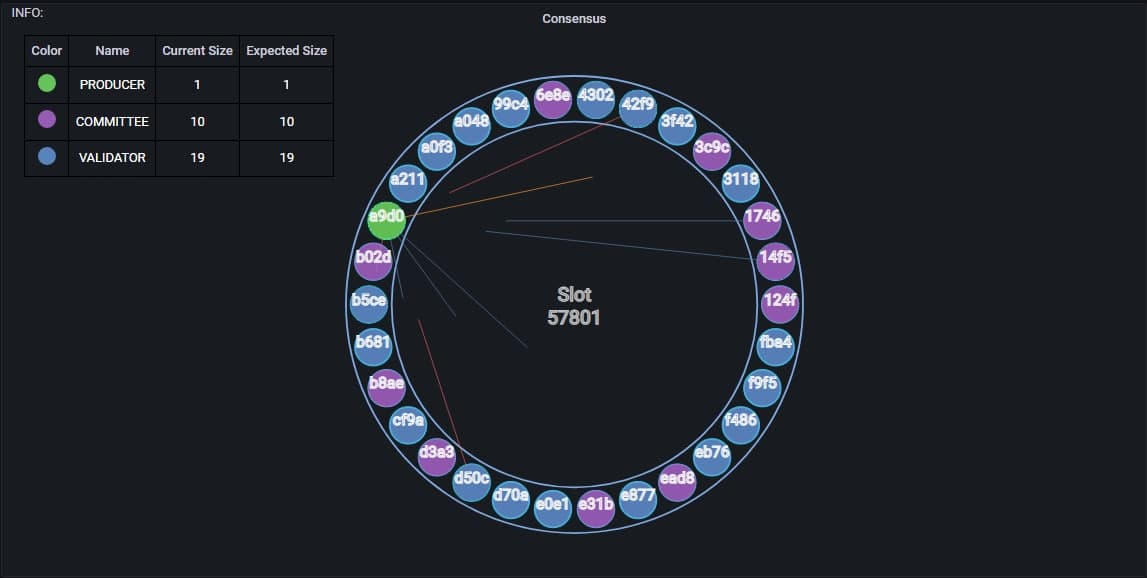}

 \caption{Consensus plugin (with legend) visualising a test network of 30 nodes in real time.}
 \label{fig:consensus}
\end{figure}

The aim of visualising the consensus mechanism is to quickly evaluate if nodes contributing to the consensus learned about their correct roles, and if they perform their roles accordingly. In order to have a scalable visualisation, nodes are placed around a circle, and scaled accordingly to the size of the network. Roles are visualized with a colour map. Each slot, nodes change their roles, and execute the protocol accordingly. To visualise the execution, the plugin visualises messages exchanged between nodes in a form of animated lines flying from an origin node to the destination node. The animations are time synchronous, and transfer times, and latencies are taken into account. Additionally, every message is logged with a type, indicating the sub protocol within which it was created. As an example, messages being sent from committee members to the block producer are attestations for the current block. The animated lines are coloured indicating the message type.
The thickness of the animate line indicates the size of the payload transferred between nodes. Figure \ref{fig:consensus} shows the consensus plugin running live visualising a test network of 30 nodes. The green coloured node indicate the block producer role for the current slot, nodes coloured violet are part of the committee, and blue nodes are validators.

\subsection{Generality}
\begin{figure}[!ht]
\centering
\includegraphics[scale=0.7]{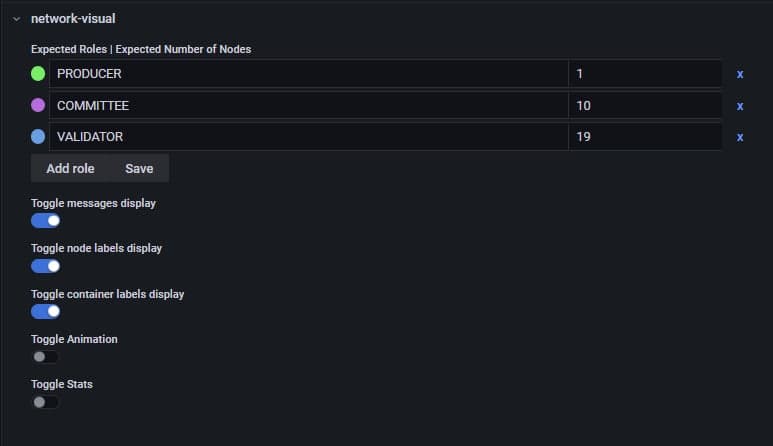}
 \caption{Consensus plugin options menu.}
 \label{fig:options}
\end{figure}

Both plugins are customizable from Grafana options menu. For example, users can add new roles, name and color them. Figure \ref{fig:options} shows the consensus plugin options menu, where users can additionally turn on or off display of messages, nodes or containers labels and so on.

\section{Acknowledgements}
The authors gratefully acknowledge the European Commission for funding the InnoRenew CoE project (H2020 Grant Agreement \#739574) and the Republic of Slovenia (Investment funding of the Republic of Slovenia and the European Union of the European Regional Development Fund) as well as the Slovenian Research Agency (ARRS) for supporting the project number J2-2504 (C).\\

\bibliographystyle{IEEEtran}
\bibliography{ref}

\end{document}